\pdfoutput=1

\documentclass[5p]{elsarticle}

\usepackage{amssymb}

\usepackage[table]{xcolor}
\usepackage{booktabs}
\usepackage{amsmath}
\usepackage{upgreek}
\usepackage{hyperref}

\journal{-}

\begin{document}

\begin{frontmatter}

\title{A forecast of the sensitivity of the DALI Experiment to Galactic
axion dark matter}

\author[label1,label2]{Juan F. Hernández Cabrera\corref{cor1}}
\ead{juan.f.hdez.cabrera@iac.es}

\author[label3,label1,label2]{Javier De Miguel\corref{cor1}}
\ead{javier.miguelhernandez@riken.jp}

\author[label1,label2]{Enrique Joven Álvarez}
\author[label1,label2]{E. Hernández-Suárez}
\author[label1,label2]{J. Alberto Rubiño-Martín\corref{cor1}}
\ead{jalberto@iac.es}

\author[label3]{Chiko Otani\corref{cor1}}
\ead{otani@riken.jp}

\cortext[cor1]{Corresponding author}



\address[label1]{Instituto de Astrof\'isica de Canarias, E-38205 La Laguna, Tenerife, Spain}
\address[label2]{Departamento de Astrof\'isica, Universidad de La Laguna, E-38206 La Laguna, Tenerife, Spain}
\address[label3]{The Institute of Physical and Chemical Research (RIKEN),
Center for Advanced Photonics, 519-1399 Aramaki-Aoba, Aoba-ku, Sendai, Miyagi 980-0845, Japan}

\begin{abstract}
The axion is a long-postulated boson that can simultaneously solve two fundamental problems of modern physics: the charge-parity symmetry problem in the strong interaction and the enigma of dark matter. In this work we estimate, by means of Monte Carlo simulations, the sensitivity of the Dark-photons$\&$Axion-Like particles Interferometer (DALI), a new-generation Fabry-Pérot haloscope proposed to probe axion dark matter in the 25--250 $\upmu$eV band.
\end{abstract}


\end{frontmatter}


\section{Introduction}
According to the current picture in cosmology, more than 4/5 of the Universe belongs to a dark sector. In  particular, dark matter, or non-luminous matter, is thought to be about five times more abundant than the baryonic matter, which represents only 1/20 of the total—e.g., see \cite{Bertone:2016nfn} for a recent review. Fortunately, despite its weak coupling to ordinary photons, we can infer the existence of dark matter through indirect channels, as noted in the pioneering work of Zwicky and, thereafter, Rubin \& Ford \cite{Zwicky:1933gu, Rubin:1970zza}. This encourages the search for dark matter comprising the Milky Way halo as a promising strategy for the first direct detection of the invisible region of the Universe.

Axions \cite{PhysRevLett.40.223}, sometimes referred to as `higgslets' \cite{PhysRevLett.40.279}, are light bosons that emerge as a consequence of the quantum chromodynamic (QCD) solution to the observed conservation of charge (C) and parity (P) symmetry in the strong interaction \cite{PhysRevLett.38.1440}, the so-called `strong CP problem,' but which, in addition, are well-motivated candidates for cold dark matter in a parameter space that has been shown to be enormously wide \cite{ABBOTT1983133, DINE1983137, PRESKILL1983127}. The axion, given a fiducial model—i.e., a Kim--Shifman--Vainshtein--Zakharov (KSVZ)-like axion does not couple to fermions in a tree-vertex \cite{PhysRevLett.43.103, SHIFMAN1980493} while, e.g., a Dine--Fischler--Srednicki--Zhitnitsky (DFSZ)-type model does \cite{DINE1981199, Zhitnitsky:1980tq}—, interacts with particles of the Standard Model. The axion-photon coupling adopts a particularly simple expression from a classical approach, which reads

\begin{equation}
\mathcal{L}_{\phi\gamma}=g_{\phi\gamma} \, \phi \, \mathord{\mathrm{E}} \cdot \mathord{\mathrm{B}}  \,,
\label{Eq_1}
\end{equation}
where $g_{\phi\gamma}$ is the coupling strength, E and B are the electric and magnetic field, respectively, and $\phi$ is the axion field.

{Numerous results constrain the axion mass ($m_{\phi}$) to coupling strength to photons parameter space compatible with axion-like particles (ALPs). Diverse experiments exclude the sector $g_{\phi\gamma}\gtrsim10^{-7}$ GeV$^{-1}$ for $m_{\phi}/$eV$\lesssim10^{-3}$ \cite{Ehret:2010mh, Betz:2013dza, DellaValle:2015xxa, OSQAR:2015qdv}. Helioscopes \cite{CAST:2017uph}, partially overlapping stellar hints based on interaction with axions in the plasma of stars \cite{Dolan:2022kul, Ayala:2014pea}, exclude $g_{\phi\gamma}\gtrsim10^{-10}$ GeV$^{-1}$ for $m_{\phi}\lesssim10^{-2}$ eV; while different astronomical campaigns and simulations rule out different sectors—c.f. \cite{Caputo:2021rux, Hamaguchi:2018oqw, Marsh:2017yvc, Meyer:2016wrm, Regis:2020fhw}. Cosmology also constrains the mass of the axion so that it does not overclose the Universe, or so that it is not excessively hot, and the search for axion in the range 10$^{-6}\lesssim m_{\phi}/$eV$\lesssim10^{-3}$ is particularly tantalizing—see \cite{Marsh:2015xka, ParticleDataGroup:2018ovx} and references therein. The two heaviest mass decades in this range are compatible with a scenario in which the axion arises after cosmic inflation \cite{Buschmann:2021sdq}. }

\begin{figure}[h]
   \centering
 \includegraphics [width=0.48\textwidth]{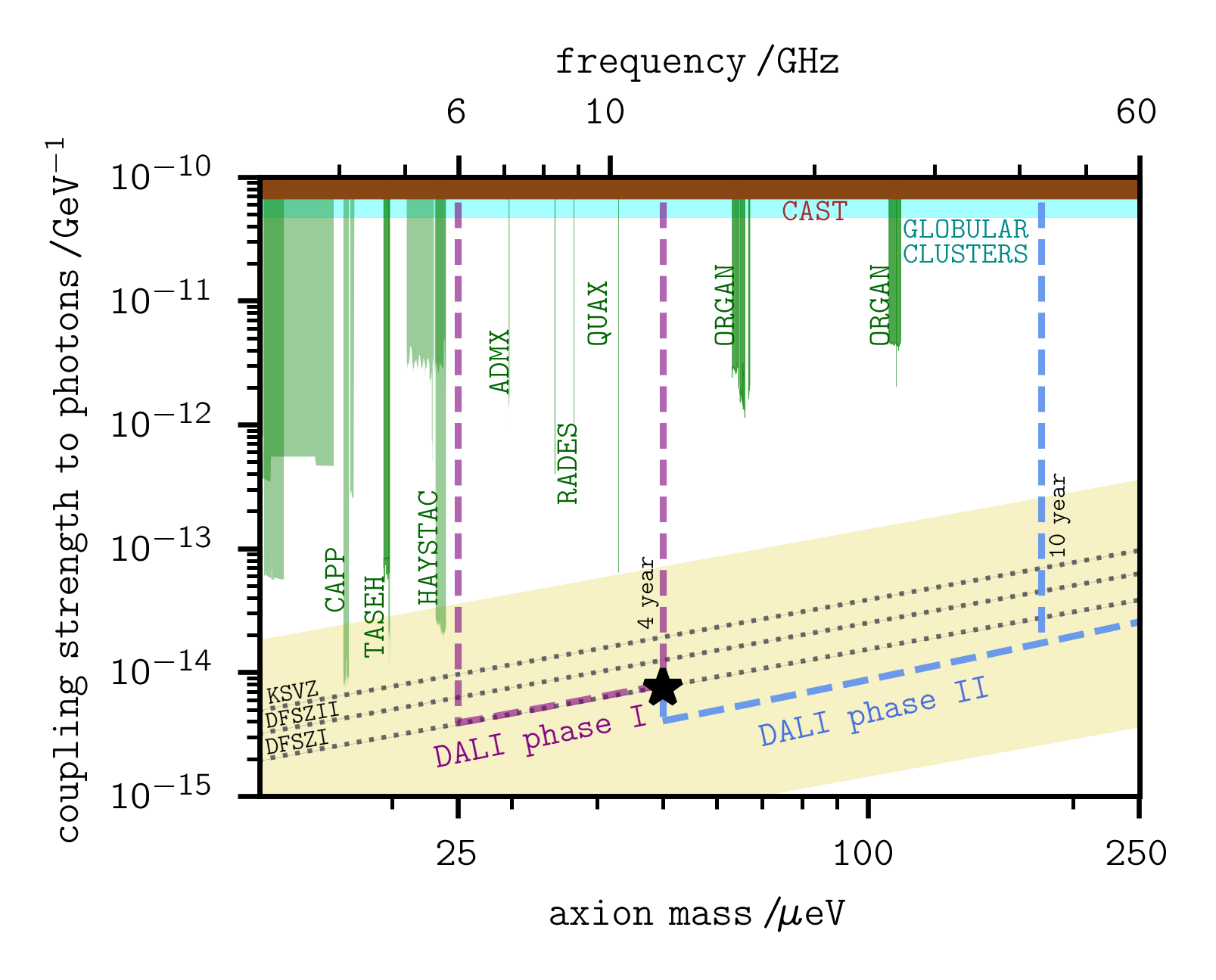}
\caption{{DALI sensitivity projection to Galactic axion dark matter with a local density of $\rho_{\phi}\sim0.45$ GeV$^{-1}$ projected onto current exclusion limits that are differentiated by color
\cite{Caputo:2021rux, Hamaguchi:2018oqw, Marsh:2017yvc, Meyer:2016wrm, Ayala:2014pea, Dolan:2022kul, Straniero:2015nvc, Regis:2020fhw, HAYSTAC:2020kwv, Ehret:2010mh, Betz:2013dza, DellaValle:2015xxa, OSQAR:2015qdv,DePanfilis:1987dk, Hagmann:1990tj, ADMX:2009iij, HAYSTAC:2018rwy, Lee:2022mnc,TASEH:2022hfm, CAST:2017uph, CAST:2020rlf, ADMX:2018ogs, Alesini:2019ajt, ADMX:2021nhd, McAllister:2017lkb, Quiskamp:2022, Quiskamp:2023ehr}. The magnetic field strength is 9.4 T for phase I, in purple, and 11.7 T for phase II, in blue \cite{Liebel2012}. The plate area is 1/2 m$^2$ and 3/2 m$^2$, respectively; and N$\sim$50 layers are stacked in series. Although the phase I configuration has a potential to probe the DFSZ I axion above 200 $\upmu$eV, phase II is introduced to accelerate the scan speed by a factor of about twenty \cite{DeMiguel:2020rpn, DeMiguel_2303.03997}. The system temperature is determined by an offset of $\sim$1 K, whose source is a physical background cryogenized with $^3$He coolers, plus $\sim$3 times the quantum noise limit at each wavelength to account for a frequency-dependent noise figure observed in high-electron-mobility transistor technology \cite{McCulloch et al., PMID:25384166}. This causes a slope with respect to the QCD models that is roughly compensated by a frequency-dependent quality factor ($Q=2\pi\nu_0\times\tau_g$), computed as $Q\sim17\times\nu_0$/GHz$\times$N, which accounts for losses and low-temperature effects on the dielectric properties of  zirconia substrates \cite{Hernandez-Cabrera:2023syh}. The instantaneous sweep bandwidth is several dozen megahertz; while the axion-induced signal linewidth is $\Delta \nu/\nu\approx5\times10^{-7}$. The KSVZ and DFSZ axion models are projected over the entire experimental range, 25--250 $\upmu$eV. The QCD axion window is shaded in yellow \cite{DiLuzio:2016sbl}. The region in white is compatible with axion-like particles (ALPs) which do not solve the strong CP problem. The star represents a DFSZ I axion with $\sim$$50$ $\upmu$eV mass.}}
\label{Fig_0}
\end{figure}

{In a standard halo model, the density of dark matter in the vicinity of the Earth is several hundred MeV$\,$cm$^{-3}$—e.g. \cite{Bertone:2016nfn}. For a typical QCD axion mass, quintillions of dark matter axions would occupy, at any time, the vessel of `haloscopes,'  ground-based detectors that search for ambient dark matter. In axion haloscopes \cite{Sikivie:1983ip}, dark matter is converted into photons via the inverse Primakoff effect by action of a static magnetic field that contributes a virtual photon—$\phi+\gamma_{\mathrm{virt}}\leftrightarrow\gamma$ \cite{Primakoff:1951iae}. Halo axions have a velocity dispersion of the order of $10^{-3}c$ and, therefore, their rest and dynamic masses approximately coincide, and $m_{\phi}\sim\omega$; $\omega$ being the angular frequency of the axion-induced radio-waves, that are scanned by a microwave receiver. Results from haloscopes in the sector of interest for this work are projected in green on Fig. \ref{Fig_0}; axion helioscopes \cite{Sikivie:1983ip} are shown in brown, and astronomical bounds to the axion in blue. The search for axion at high frequency remains poorly explored. A reason behind this is that in the most widespread haloscope concept, the frequency at which the faint signal originating from axion-to-photon conversion is enhanced scales with the diameter of a closed resonant cavity, or the distance between the mechanisms inserted within it to allow for tuning, as $d\sim c/\nu_0$; $c$ being the speed of light and $\nu_0$ the resonance frequency. A progressive reduction of $d$ at higher frequencies contracts the magnetized volume in which dark matter axions oscillate into photons, lowering the power output in the form $P \propto V$. Together with a series of hurdles that accompany the miniaturization of the devices, this gradually reduces the scan speed and sensitivity of cavity-haloscopes probing heavier axions—e.g., see \cite{Stern:2015kzo}. }
\begin{figure}[h]
   \centering
 \includegraphics [width=0.4\textwidth]{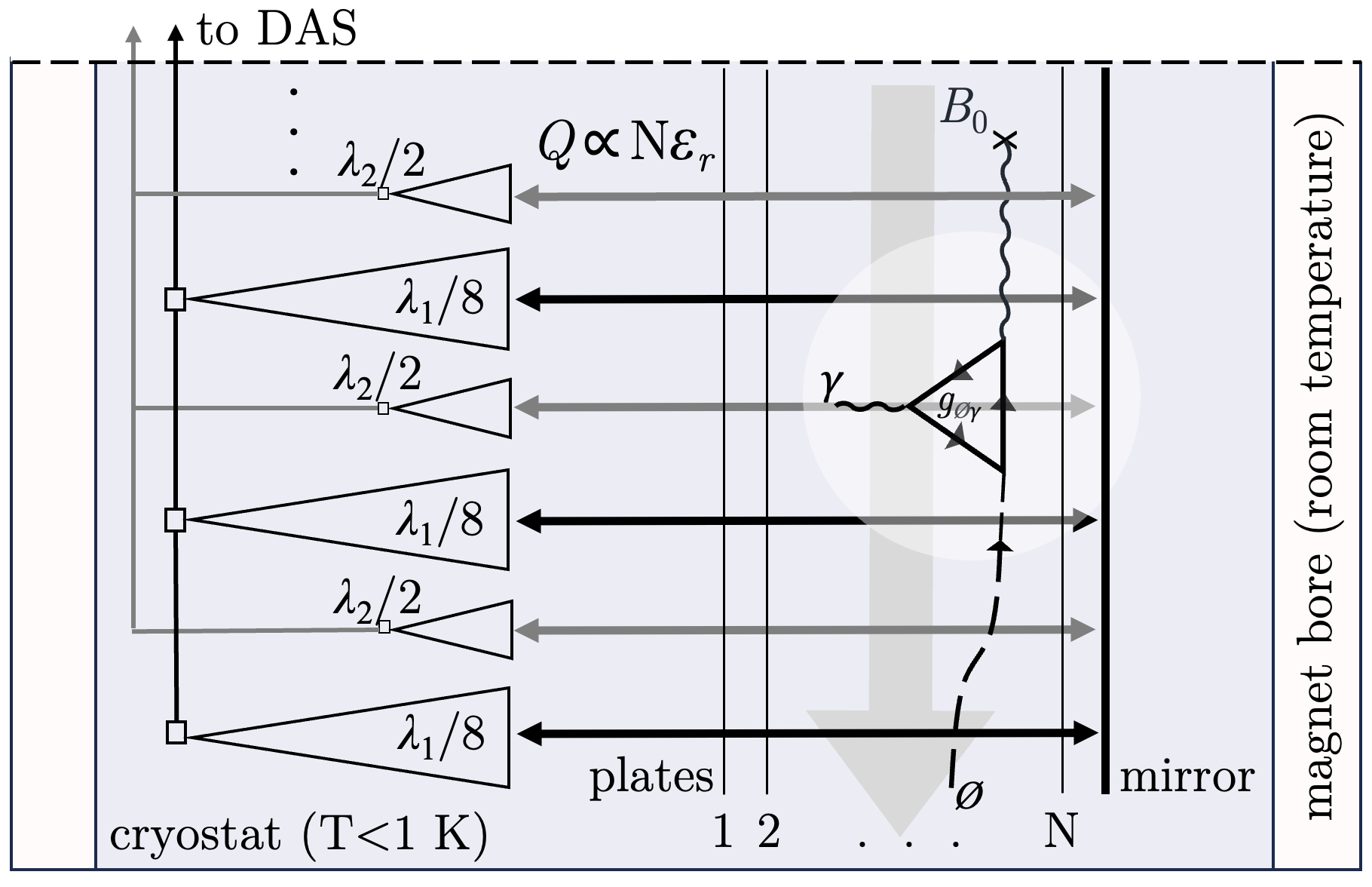}
\caption{{Schematic diagram of DALI. The experiment cryostat is housed within the bore of a regular superconducting magnet \cite{Liebel2012}. Microwaves arise in an axion-photon-photon vertex. The resonator is composed of dielectric boundaries, with permittivity $\varepsilon_r$, and a top mirror. The quality factor ($Q$) is a figure of merit of signal enhancement around a center frequency. The resonant frequency at which the axion is probed is tuned by setting a distance of a fraction of the wavelength between the N+1 boundaries by means of non-magnetic electromecanical actuators, not shown. The resonance is observed simultaneously at two frequencies about four wavelengths apart, $\sim$$\lambda_1/2$---higher frequency---and $\sim$$\lambda_2/8$---lower frequency---; $\lambda$ being the wavelength of the photon originating from axion conversion \cite{DeMiguel_2303.03997}. The phased antenna array  receives both harmonics simultaneously, by means of different pixels, and data are stored by independent data acquisition systems (DAS). Channels of equal frequency are combined using methods similar to radio interferometry that allow for post-processing with a negligible error \cite{2010A&A...520A...4B, Rubino-Martin:2023fya}.   }}
\label{fig_00}
\end{figure}

{The Dark-photons \& Axion-Like particles Interferometer (DALI) is a proposal for the search of Galactic dark matter in a mass range above a few dozen microelectronvolts. To overcome some of the challenges encountered by high-frequency cavity-haloscopes operating in this axion mass range, DALI replaces the closed cavity used in the cavity-haloscope, or the parallel superconducting wires as addressed in the Sikivie Fabry-Perot haloscope concept \cite{Sikivie:1983ip, Sikivie:1993jm, Rybka:2014cya}, with a dielectric-layer Fabry-Pérot resonator. In a Fabry--Pérot interferometer, the resonant frequency is tuned by setting a wavelength fraction of distance between adjacent plates. Since the frequency tuning is uncoupled to the plate area, and the power scales with its size as $P \propto A$, this enables access to axions with shorter Compton wavelengths with a greater sensitivity, opening the door to the exploration of the cosmic axion in a post-inflationary scenario \cite{Buschmann:2021sdq}. Different collaborations, such as MADMAX \cite{Majorovits:2016yvk, MADMAX:2019pub}, ADMX-Orpheus \cite{Cervantes:2022epl, Cervantes:2022yzp}, LAMPOST \cite{Baryakhtar:2018doz, Chiles:2021gxk}, MuDHI \cite{Manenti:2021whp} and DBAS \cite{McAllister:2017ern, Quiskamp:2020yrx}, bring similar approaches for axion quests in a parameter space that is vast enough to accommodate multiple experiments. New features introduced by DALI are an altazimuth platform to enhance its sensitivity to dark matter substructures that may be cruising the Solar System—cf. \cite{OHare:2017yze, Knirck:2018knd, OHare:2018trr, Tinyakov_2016, Berezinsky_2013, PhysRevD.97.083502}—; or a multi-pixel focal plane where the signals are combined in a post-processing that relies on methods which we originally developed for radio astronomy, and which allows for larger plate areas and, potentially, the simultaneous scanning of multiple axion masses separated by a spectral distance of about four wavelengths—cf. \cite{DeMiguel:2020rpn, DeMiguel_2303.03997} for the details. This multi-frequency approach is inspired by previous cosmic microwave background experiments—e.g., \cite{2010A&A...520A...4B, Rubino-Martin:2023fya}—and, once implemented, could double the scan speed of the haloscope over a broad experimental range. A scheme of the work principle is shown in Fig. \ref{fig_00}. On the other hand, the DALI design has been focused on exclusively using commercially available hardware to allow for a fast, cost-effective and simple manufacture which eliminates the need for a previous stage of technological development. The experiment, to be installed at the Teide Observatory, in the Canary Islands—longitude 16º 30' 35'' West; latitude 28º 18' 00'' North—, is currently in a design and prototyping stage.}

{A sensitivity projection of DALI on the mass-coupling plane is shown in Fig. \ref{Fig_0}. Here, the black star represents a DFSZ I axion of 50 $\upmu$eV mass. The prefactor in the first term of the right hand side of Eq. \ref{Eq_0} is to be verified in this article by means of Monte Carlo simulations in which this synthetic axion signal—the star—is injected on a background with noise,}
\begin{equation}
\begin{aligned}
g_{\phi \gamma}\!\gtrsim\!2.7\times10^{-13} \,\mathrm{GeV^{-1}} \times \left(\frac{\mathrm{SNR}}{Q}\right)^{1/2}  \!\! \times \left(\frac{\mathrm{m}^2}{A}\right)^{1/2}\!\! \times  \\\left(\frac{m_{\phi}}{\mathrm{\upmu eV}}\right)^{5/4}
 \!\! \times \left(\frac{1\,\mathrm{s}}{t}\right)^{1/4} \!\! \times \left(\frac{T_\mathrm{{sys}}}{\mathrm{K}}\right)^{1/2}  \!\! \times \frac{\mathrm{1\,T}}{B_0} \times \left(\frac{\mathrm {GeV cm^{-3}}}{\rho_{\phi}}\right)^{1/2} \, .
\end{aligned}
\label{Eq_0}
\end{equation}
{The operational principle of a Fabry--Pérot interferometer such as DALI is based on the constructive interference between incoming and reflected waves that are transparent to the dielectric boundaries, giving rise to a standing wave pattern. The output is spectrally modified compared to the input beam, allowing for power enhancement in relatively narrow frequency bands centered at a tunable resonant frequency \cite{1899ApJ.....9...87P}. Focusing on the parameters over which one has control, its sensitivity to axion dark matter scales as $g_{\phi\gamma} \propto (T_{\mathrm{sys}}\,  A \, Q)^{-1/2} \, t^{-1/4} \, B^{-1}_0$; with $T_{\mathrm{sys}}$ the system temperature, $A$ the cross-sectional area, $Q$ the quality factor, $t$ the integration time—typically $t \lesssim 1$ ms per sub-spectra, to mitigate 1/f noise, which are stacked into a single spectrum at each frequency to mitigate white noise \cite{doi:10.1063/1.1770483}—, and $B_0$ is the magnetic field strength—in Eq. \ref{Eq_0}, SNR is signal-to-noise ratio and $\rho_{\phi}$ is the local density of axion dark matter. Therefore, in addition to procuring a powerful superconducting solenoid with a large bore which allows for housing a low-temperature cryostat with a large cross-sectional area—e.g. \cite{Liebel2012}—, it becomes crucial to provide a high $Q$ factor. The quality factor of a Fabry--Pérot resonator, defined as 2$\pi$ times the ratio of the energy stored in the resonator and the energy loss per oscillation period, is $Q=\omega \tau_g$, where the group delay time, $\tau_g$, is the average lifetime of a photon in the resonator or, equivalently, decay time of the energy density of radiation in the interferometer, and adopts typical values of a few nanoseconds per layer \cite{Renk}. The $Q$ factor scales with the number of layers in series, and a higher relative permittivity, $\varepsilon_r$, results in a higher peak and a narrower full width at half maximum (FWHM) of the Lorentzian spectral feature caused by interferometry. A benchmark instantaneous scanning bandwidth of $\sim$50 MHz, where the quality factor $Q\!\sim\!10^{4}$ needed to reach QCD axion sensitivity is tenable in practice for about four dozen layers made of zirconia, is compatible with our experimental results \cite{Hernandez-Cabrera:2023syh}.
}

In this manuscript we report the simulation of an observational campaign intended to find the axion in the vicinity of the black star shown in Fig. \ref{Fig_0}—a DFSZ I axion of $\approx$50 $\upmu$eV mass. The rest of the manuscript is structured as follows. In Sec. \ref{II} we define the statistical model, the data processing is described in Sec. \ref{III}, while the results are reported and discussed in Sec. \ref{IV}.

\section{Methods}\label{II}

\subsection{Data analysis} \label{Data_Analysis}

A data analysis method has been envisioned for DALI to detect and quantify local power excesses. As a methodology designed for a new-generation Fabry-Pérot axion haloscope, it brings the statistical methods used in resonant cavity haloscopes \cite{Brubaker:2017rna, Quiskamp:2022} closer to its architecture. The goal of this procedure is to flag all candidates in the spectrum representing a power excess plausibly caused by an axion with a given confidence level. During the data acquisition, a set of $M$ partially overlapping spectra with a size of $2^N$ bins each will be obtained by tuning the resonator at different frequencies. The use of a power of two for the size of the spectra is intended to minimize the computing time of fast Fourier transform (FFT) algorithms \cite{Gambron2020}. Each spectrum will be the result of averaging power data in a $\sim 50$ MHz frequency range during an effective time period $t$.

Arrays of $2^N$ time domain samples will be normalized in power before calculating their FFTs individually and averaging out the resulting $2^N$-bin power spectra. $N$ will be chosen to limit the duration of the recorded signal to $2^N f_s \sim$1 ms to mitigate gain fluctuations given typical values in regular radiometers \cite{Gallego2004}. This will unvariably create a lower bound for the bin width $\Delta \nu_b = 1 / (2^N f_s)$. 

Additional data---e.g. resonant frequency stability---may be used to flag a subspectrum as compromised and therefore discard it.

Intermediate frequency (IF) interferences existing in a downconverted spectrum are expected to occur at approximately the same location in each of the IF spectra. After removing compromised spectra, the procedure for eliminating these interferences involves dividing the average spectrum by the estimated spectral baseline, which is obtained by means of a Savitzky-Golay (SG) filter. Bins with a power excess or defect equal or greater than an arbitrary threshold are identified as potentially affected by IF interferences and iteratively substituted by random numbers generated according to a Gaussian distribution with the same mean and standard deviation as the normalized average spectrum until no further such bins are found. The same procedure is also applied to the three neighboring bins of each instance in each side. In addition, the same bin indices where such interferences were identified in the normalized average spectrum are also substituted by randomly generated values in individual spectra in the same manner.

Radio frequency (RF) interferences are flagged in the individual spectra resulting from the previous procedure as all bins with a deviation exceeding a threshold based on the exclusion sector to be analyzed, and are iteratively substituted by randomly generated numbers until no further instances of such bins remain present. The resulting spectra are then normalized to their individual baselines, which are estimated by means of a SG filter, resulting in a set of processed spectra. The bins of the processed spectra will follow a Gaussian distribution with mean $\mu^p = 0$ and standard deviation $\sigma^p = 1 / \sqrt{t \Delta \nu_b}$.

Even though the bins potentially affected by IF or RF interferences are replaced by random values to eliminate their effect on the outcome of this method, their indices are listed so as to exclude them from the scanned region. 

A window length of $W$ = 3001 and a polynomial degree of $d$ = 2 has been used for all steps involving SG filters in this simulation. Larger values of $d$ and lower values of $W$ result in a reduction of the filter cutoff frequency, which has been prioritized, since filter-induced rippling around artificial axion signals has been observed in average spectra obtained in Monte Carlo simulations implementing filters with higher cutoff frequencies. These parameters may be modified, however, when different data are analyzed depending on the complexity of the baseline and the spectrum length. Filters with better stopband attenuation have been considered—e.g. modified sinc, weighted SG and others \cite{Schmid2022}—but ruled out because of their poor transition band behavior compared to the SG filter. The effect of the imperfect stopband behavior of the SG filter in this application will be observed as nonzero correlations among neighboring bins. 

\begin{figure*}
    \centering
    \includegraphics[width=0.8\textwidth]{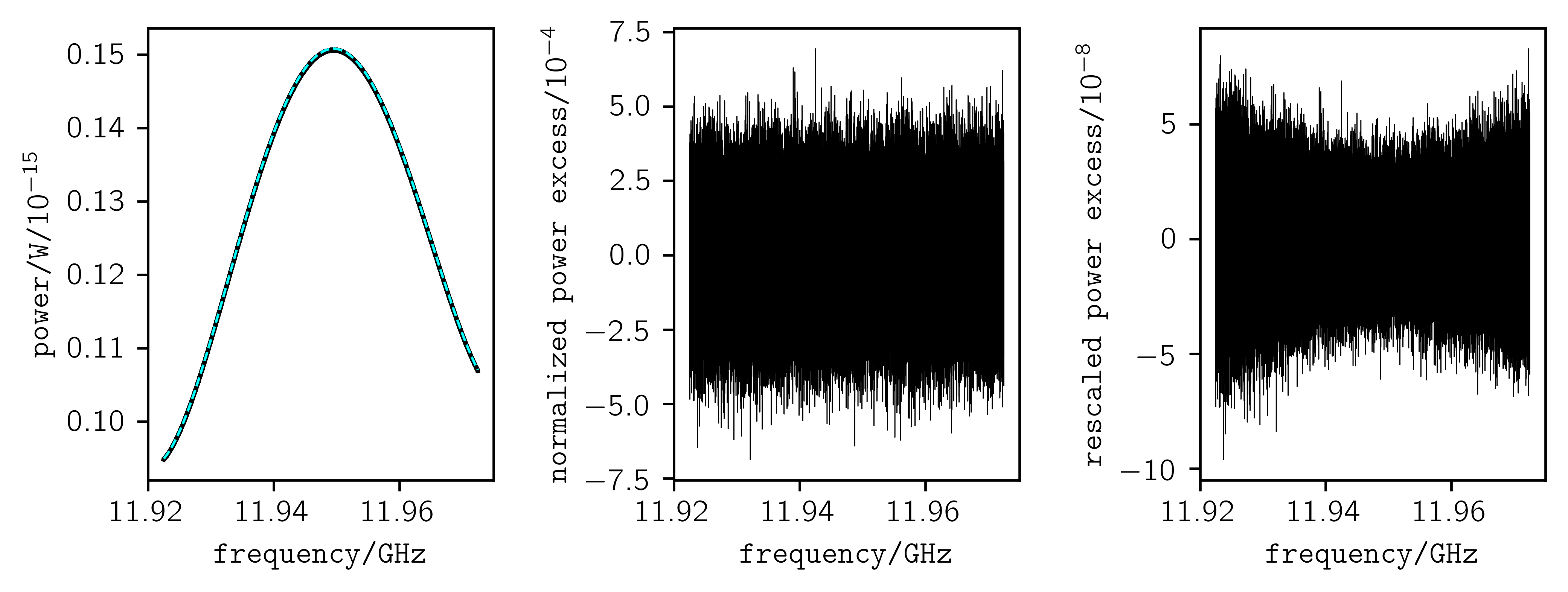}
    \caption{A representation of a numerically generated spectrum at different points of the analysis procedure. {The spectrum baseline has been obtained by simulation.} Left: raw spectrum (black) and the estimated baseline obtained by means of a Savitzky-Golay filter ({light blue}). Center: processed spectrum. Note that the normalization in this case refers to the estimated baseline. Right: Rescaled spectrum. }
    \label{Fig_1}
\end{figure*}

Given that a lower enhancement is expected further away from the resonating frequency, the processed spectra $\delta_{ij}^p$ and their standard deviation $\sigma_{ij}^p$ are rescaled bin-wise by multiplying their values by the inverse of the expected quality factor at each bin $Q_{ij}$, namely

\begin{equation}
    \delta_{ij}^s = \frac{\delta_{ij}^p}{Q_{ij}} \, ,
    \label{Eq_2}
\end{equation}

to yield the rescaled spectrum bins $\delta_{ij}^s$, where $i$ and $j$ represent the spectrum index and the bin index, respectively. The rescaled standard deviations $\sigma_{ij}^s$ are obtained in the same manner. The resonator output is expected to have a Lorentzian-like transfer function of the form $Q (\nu) \sim K / (1 + [2 (\nu - \nu_0) / \mathrm{FWHM}]^2)$. The aim of rescaling the spectra is not to assign specific dimensional meaning to the rescaled values, so the rescaled bins may be regarded as having arbitrary dimensions. As a result of this operation, bins will cease to be random values drawn from Gaussian distributions with the same standard deviation, and therefore individual values of $\sigma$ are henceforth considered for each bin—see Fig. \ref{Fig_1}.

The resulting rescaled spectra are now merged in a single stacked spectrum. IF bins in non-overlapping bands map to a single RF bin, and their values are copied directly to the corresponding RF bin in the stacked spectrum. Otherwise, all IF bins corresponding to a RF bin in overlapping bands are combined by using a maximum-likelihood (ML) sum with weights $w_{ijk}$ given by

\begin{equation}
    w_{ijk} = \frac{\left (\sigma_{ij}^s \right )^{-2}}{\sum_{i'} \sum_{j'} \left (\sigma_{i' j'}^s \right )^{-2}} \, .
    \label{Eq_3}
  \end{equation}

The bin index in the stacked spectrum is represented by $k$. The sum range in Eq. \ref{Eq_3} is to be interpreted as all bins with indices $i$ and $j$ mapping to a given RF bin with index $k$. The stacked spectrum bins in overlapping bands $\delta_k^c$ will be computed from rescaled bins $\delta_{ij}^s$ as

\begin{equation}
    \delta_k^c =  \sum_{i'} \sum_{j'} w_{ijk} \delta_{ij}^s \,,
    \label{Eq_4}
\end{equation}

where the sum range should be interpreted in the same way as in Eq. \ref{Eq_3}. Individual bins may be considered to be random values drawn from independent Gaussian distributions if the effect of the SG filtering is ignored. Therefore, the variance of the weighted sum of bins is equal to the weighted sum of variances with squared weights, and consequently the standard variation of the stacked spectrum bins is
    
\begin{equation}
    \sigma_k^c =  \sqrt{\sum_{i'} \sum_{j'} w_{ijk}^2 \left ( \delta_{ij}^s \right )^2} \, .
    \label{Eq_5}
\end{equation}

Further numeric manipulation of the data must take into account that the axion-induced signal will be present in more than one consecutive bin. The subsequent steps in this procedure will adapt this manipulation to such a signal with a width larger than $\Delta \nu_b$ by means of a ML sum that will maximize the SNR for an axion-induced signal whose power excess is concentrated in the $\ell$th bin of the resulting `grand' spectrum. First, a `rebinned' spectrum is obtained by combining non-overlapping segments of the stacked spectrum of length $K_r$. If $D_k^c = \delta_k^c (\sigma_k^c)^{-2}$ and $R_k^c = (\sigma_k^c)^{-1}$, this rebinning is computed as

\begin{equation}
     D_\ell^r = \sum_{k = (\ell - 1) K_r + 1}^{k = \ell K_r} D_k^c \,
     \label{Eq_6}
\end{equation}

and

\begin{equation}
    R_\ell^r = \sqrt{\sum_{k = (\ell - 1) K_r + 1}^{k = \ell K_r} (R_k^c)^2} \, .
    \label{Eq_7}
\end{equation}

Then, a weighted ML sum is calculated across overlapping segments of the rebinned spectrum with a length of $K_g$ bins, resulting in the previously mentioned grand spectrum. In this implementation, we have found that an optimal choice of $K_r$ and $K_g$ is such that $K_r K_g$ bins are approximately 1.5 times as wide as the axion FWHM. The weights of this sum must take into consideration the expected axion line shape as a consequence of a space-domain Maxwell-Boltzmann distribution \cite{Brubaker:2017rna}, namely
    
\begin{equation}
    S_\phi \left ( \nu \right ) 
    = K \sqrt{\nu - \nu_\phi} \exp{\left (- \frac{3 (\nu - \nu_\phi)}{\nu_\phi \frac{\langle v_\phi^2 \rangle}{c^2}} \right )} \,,
    \label{Eq_8}
\end{equation}

where $K$ is a constant to scale the axion spectrum such that the total axion signal power $P_\phi = \int_{-\infty}^{\infty} S_\phi(\nu) \mathrm{d} \nu$ equals the expected quantity \cite{Sikivie:1983ip}. It can be shown that Eq. \ref{Eq_8} implies an axion FWHM equal to
    
\begin{equation}
    \begin{split}
        \mathrm{FWHM}_\phi = \frac{1}{6} \frac{\langle v_\phi^2 \rangle}{c^2} 
        \left [ W_0 \left (- \frac{1}{4e} \right ) - W_{-1} \left (- \frac{1}{4e} \right ) \right ] \nu_\phi \\
        \approx 4.848 \cdot 10^{-7} \nu_\phi \,,
    \end{split}
    \label{Eq_9}
\end{equation}

where the function $W_b$ is the $b$th branch of the Lambert $W$ function. By defining $\sigma_\ell^r = (R_\ell^r)^{-1}$ and $\delta_\ell^r = D_\ell^r (\sigma_\ell^r)^{2}$, it can be shown that each bin $\delta_\ell^r$ of the rebinned spectrum is a random value drawn from an independent Gaussian distribution whose mean will be nonzero in the case that a power excess is present, and zero otherwise, in the absence of filter-induced correlations. The standard deviation will be $\sigma_\ell^r$ regardless of the presence of a power excess.

The weights to calculate the grand spectrum can be obtained by integrating $S_\phi(\nu)$ over portions with the width of a rebinned spectrum bin; hence we define for all $q = 1, \ldots, K_g$ 

\begin{equation}
    L_q \left ( \delta \nu_r \right ) =
    K_g \int_{\nu_\phi + \delta \nu_r + (q-1) K_r \Delta \nu_b}^{\nu_\phi + \delta \nu_r + q K_r \Delta \nu_b} S_\phi(\nu) \mathrm{d} \nu
    \label{Eq_10}
\end{equation}

as a measure of the axion power contained in each rebinned bin given a misalignment $\delta \nu_r$, which can take any value in the interval $-z K_r \Delta \nu_b \le \delta \nu_r \le (1-z) K_r \Delta \nu_b$ for $0 \le z \le 1$. The weights $\bar{L}_q$ are calculated by taking the average of $L_q \left ( \delta \nu_r \right )$ over all defined values of $\delta \nu_r$. This interval and ultimately the values of the weights depend on the choice of $z$, which is selected so that the value of the sum $\sum_q L_q / K_g$ is larger than what would be obtained if the limits of the sum were shifted by 1 up or down for all values of $\delta \nu_r$.

The grand spectrum is obtained as

\begin{equation}
    D_\ell^g = \sum_q D_{\ell + q - 1}^r \bar{L}_q
    \label{Eq_11}
\end{equation}

and

\begin{equation}
    R_\ell^g = \sqrt{ \sum_q \left ( R_{\ell + q - 1}^r \bar{L}_q \right )^2} \, .
    \label{Eq_12}
\end{equation}

Now, the normalized grand spectrum is calculated as $\delta_\ell^g / \sigma_\ell^g = D_\ell^g / R_\ell^g$. In the absence of correlations induced by the SG filter, the bins of the normalized spectrum can be considered random values drawn from independent Gaussian distributions with a standard deviation equal to 1 and a nonzero mean in the absence of a power excess. In the presence of an axion-induced power excess in a given bin, the mean of its Gaussian distribution will instead be equal to the axion SNR after being reduced by the non-ideal behavior of the SG filter, since these correlations reduce the width of the histogram and cause the normalized grand spectrum to have a standard deviation $\xi < 1$. This effect can be accounted for by calculating a corrected standard deviation $\tilde{\sigma}_\ell^g = {\sigma}_\ell^g / \xi$; a corrected normalized grand spectrum can now be calculated as $\delta_\ell^g / \tilde{\sigma}_\ell^g$.

\subsection{Sensitivity estimation} \label{Sensitivity_Estimation}

The described procedure has been tested by means of a Monte Carlo simulation with two goals: testing the functionality of the procedure itself and comparing the statistical properties of the output against the naive estimate from refs. \cite{DeMiguel:2020rpn,  DeMiguel_2303.03997}. A set of power spectra exhibiting the expected physical properties resulting from the experimental setup has been simulated and an DFSZ I axion \cite{DINE1981199, Zhitnitsky:1980tq} with a line shape given by Eq. \ref{Eq_8} has been injected.

In this work we report the result of a Monte Carlo simulation after computing $\sim$20,000 iterations in a haloscope with a system noise temperature given by, roughly, the sum of the physical temperature of the amplifiers, about 1 K, plus three times the quantum noise level—which is compatible with the noise figure of commercially available amplifiers—and a quality factor $Q\simeq 10^4$ at 50 $\upmu$eV. For each spectrum, the input of the analysis procedure has been obtained as random values according to a Gaussian distribution with with mean $\mu = k_{\mathrm{B}} T_\mathrm{{sys}} \Delta \nu_b$ and standard deviation $\sigma = k_{\mathrm{B}} T_\mathrm{{sys}} \sqrt{\Delta \nu_b / t }$ \cite{doi:10.1063/1.1770483} multiplied by the quality factor curve of the Fabry-Pérot resonator. This procedure may sometimes locate an axion in a bin different from the expected one. In order to overcome this limitation, a 300-bin long interval including the expected location of the axion is recorded in each iteration and averaged out. The bin containing most axion peaks can be easily found as the bin with the greatest average value in the interval. The instantaneous bandwidth considered in this simulation is 50 MHz and the spectrum length is $2^{17}$ $(\approx 1.3 \times 10^5)$ bins. An artificial axion signal has been injected as $S_{a, \,ij} \approx \Delta \nu_b S_a (\nu_{i j})$ and has been flagged and quantified. Considering $\langle v_\phi^2 \rangle = $ (270 km/s$)^2$ in Eq. \ref{Eq_9}, an axion signal at this frequency spans approximately 23 bins; consequently, we have set $K_r =$ 6 and $K_g = 4$. In addition, $z = 0.63$ has been found to meet the requirements described in Sec. \ref{Data_Analysis}. Although the axion width will vary throughout the scan band, given its low bandwidth---$M = 2$ spectra with an overlap of 5 MHz---, this variation is expected to be negligible.

\begin{figure}[h]
   \centering
 \includegraphics [width=0.4\textwidth]{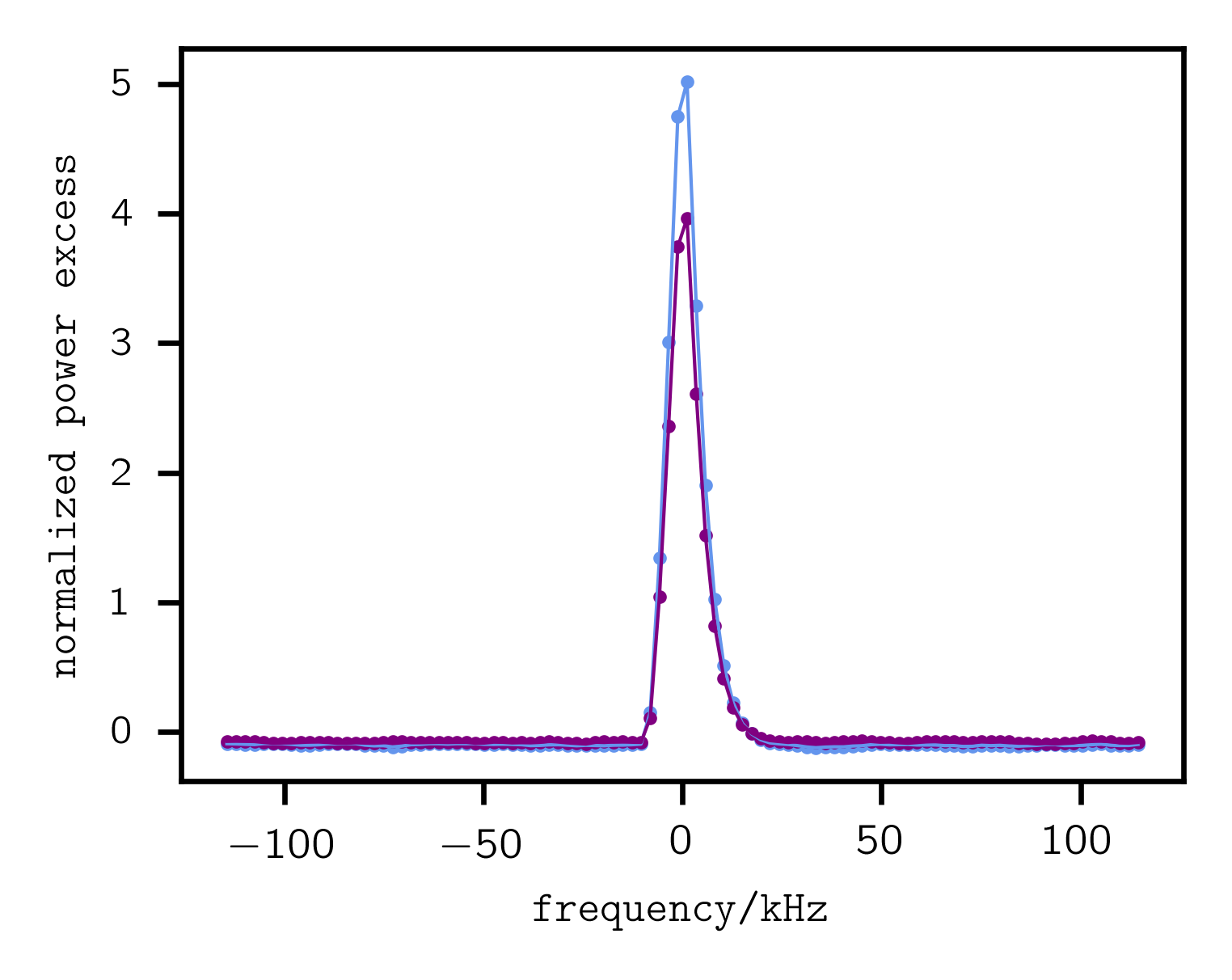}
\caption{Detection significance. The average of the output of the analysis procedure---normalized grand spectrum---computed in $\sim$20,000 iterations is represented in a 300-bin window. The peak value represents the SNR of the axion signal. {Purple: DALI phase I with $A\sim1/2$ m${^2}$; $B_0\sim 9.4$ T; $t\sim$14 day; $Q\sim10^4$. Blue: DALI phase II with $A\sim3/2$ m${^2}$; $B_0\sim 11.7$ T; $t\sim$1.3 day.}}
\label{Fig_2}
\end{figure}

\begin{figure}[h]
   \centering
 \includegraphics [width=0.4\textwidth]{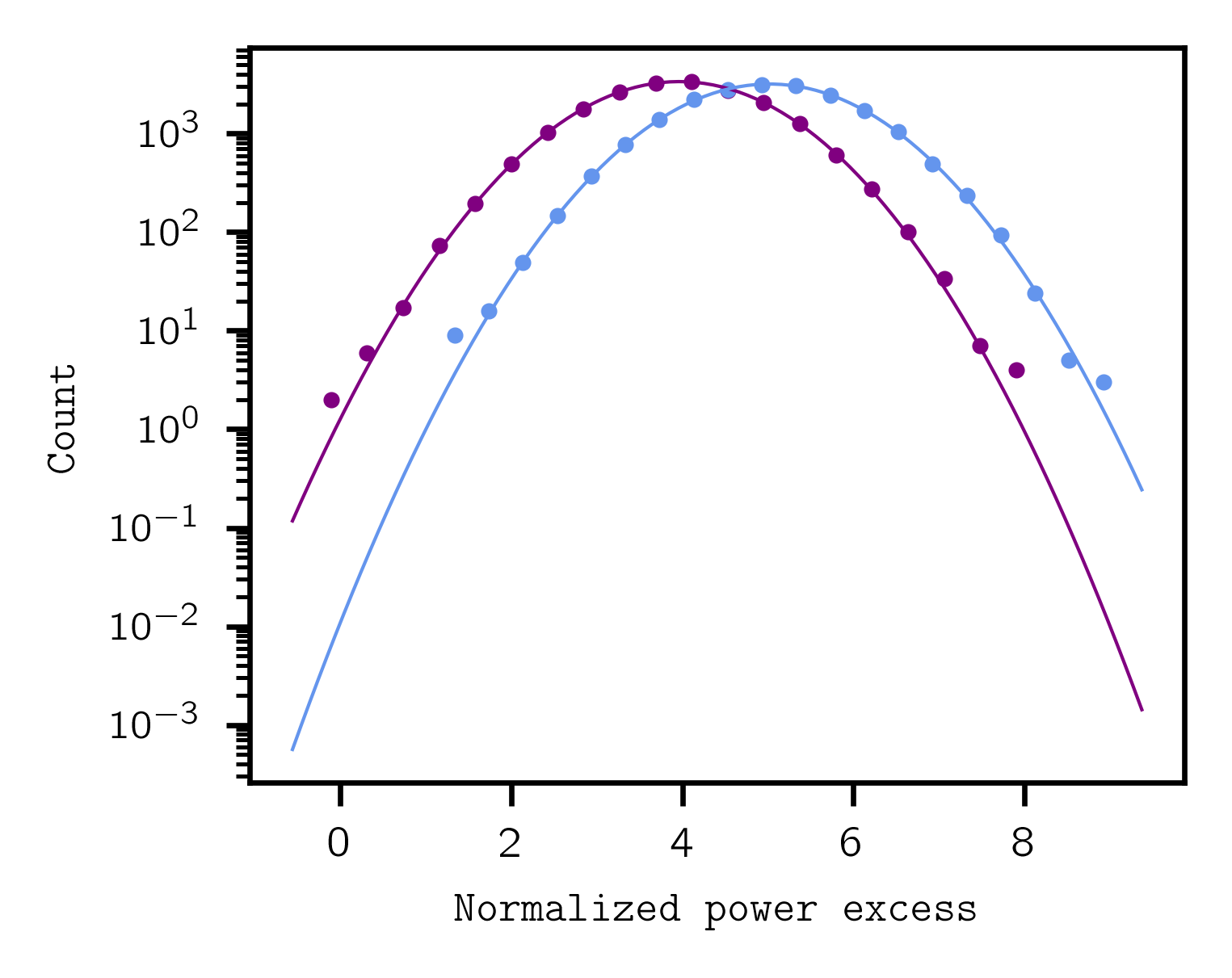}
\caption{A histogram of the normalized power excess of a synthetic axion signal as computed in $\sim$20,000 iterations. {Purple: DALI phase I with $A\sim1/2$ m${^2}$; $B_0\sim 9.4$ T; $t\sim$14 day; $Q\sim10^4$. Blue:  DALI phase II with $A\sim3/2$ m${^2}$; $B_0\sim 11.7$ T; $t\sim$1.3 day. A mean of 3.97 and a standard deviation of 0.997 for phase I; and a mean of 5.02 and a standard deviation of 0.999 for phase II}, are found by fitting the data to a continuous Gaussian distribution, represented here with a continuous line. The mean of this distribution corresponds to the peak value of the curve in Fig. \ref{Fig_2}. These results match the theoretically expected SNR from Eq. \ref{Eq_0}}.
\label{Fig_3}
\end{figure}

\subsection{Candidate selection} \label{Candidate_Selection}

The candidate selection criteria must be planned out as a trade-off between the percentage of candidates that must be rescanned and the sensitivity of the instrument. An exclusion threshold $\Theta$ is established with an arbitrary confidence level, CL, as $\Theta = \mathrm{SNR} - \Phi^{-1} (\mathrm{CL})$, where $\Phi$ is the cumulative distribution function of the standard normal distribution. The choice of $\Theta$ will determine the number of candidates $\hat{S}$ that can be expected to be flagged in any scan according to 

\begin{equation}
    \hat{S} = N_g \left [1 - \Phi(\Theta) \right ] \, ,
    \label{Eq_13}
\end{equation}

where $N_g$ stands for the number of bins in the grand spectrum, since there is a nonzero probability of the value of a given bin having a power excess larger than $\Theta$ according to its Gaussian distribution by pure chance, therefore appearing as a false positive. The choice of $\Theta$ will entail flagging as candidate an axion with a given SNR with a probability equal to CL. The proportion of bins above the exclusion line---therefore false negatives---can be reduced by selecting a larger value of CL at the expense of also increasing $\hat{S}$.

Given the statistical significance at which a hypothetical axion signal is expected to be detected as obtained in the Monte Carlo simulations reported in Sec. \ref{Sensitivity_Estimation}, the thresholds are set to $\Theta = 2.33$ and $\Theta = 3.38$ for phase I and phase II, respectively, at CL $= 0.95$. A result, percentages of rescan candidates $\hat{S} / N_g$ of $ 1.00 \%$ and $ 0.04 \%$ are expected.

\section{Data processing} \label{III}

The final result of the data acquisition campaign will be a high-resolution, low-dispersion estimation of the power spectral density function of a microwave signal, which is obtained by calculating the discrete power spectrum of the time domain samples in $L$ segments of $2^N$ samples, normalizing each segment to its measured power and averaging out the frequency-domain power segments. This can be expressed using a continuous time formalism as

\begin{equation}
   S(\nu) \simeq \frac{1}{L} \sum_{i = 0}^L \frac{| \mathcal{F} \{ f_i (t) \}|^2}{\frac{1}{T} \int_0^T f_i^2(t) \mathrm{d} t} \, .
   \label{Eq_14}
\end{equation}

The non-linear nature of this sum implies that the FFT must be calculated for each time-domain segment. In practical terms, the frequency domain data can be obtained in two ways: by saving time-domain data and calculating the FFTs in segments of $2^N$ samples and their average as a post-processing step or by computing FFTs as the samples are generated in $2^N$-sized segments and averaging them out, all of this online. The high sampling rate necessary to acquire the signal and the long integration period would require immense storage capabilities if the former option were implemented. It should also be considered that the time-domain samples are not relevant for the purposes of this work after the spectra have been computed. Consequently, Eq. \ref{Eq_14} will be implemented as an online, discrete-time, segment-wise FFT algorithm in a field programmable gate array (FPGA) board which will also average the frequency-domain spectra.

\section{Results and discussion}\label{IV}
DALI is a proposal for a new wavelike dark matter haloscope. In addition to a high sensitivity to Galactic axions and paraphotons, enhanced by a Fabry-Pérot interferometer tunable over broad bands, the apparatus has been designed to employ available equipment, which reduces its overall cost, the maintenance required; and the research and development effort, thus giving readiness to the experiment. In particular, DALI has been designed to employ a regular solenoid or multicoil type superconducting magnet, identical to those used in magnetic resonance imaging (MRI) over decades. The DALI project brings some additional novelties, such as a steerable platform intended to improve its directional sensitivity, which could be useful in the search for dark matter substructures navigating the Solar System. One important aspect about directionality is that both scenarios, an isotropic Halo or under the premise of an axion flow, are always simultaneously explored, so that the experiment does not lose any capability but adds functionality. Other interesting advantage of DALI is the capacity to probe two different resonance frequencies, spaced approximately four wavelengths apart, simultaneously, by means of a multi-frequency phased array of receivers housed in the same cryostat of the tuner. This can double the scan speed over a broad band. The experimental set-up and multi-channel data post-processing are inspired by previous radio telescopes for cosmic microwave background observations to which we have contributed. The DALI project is planned in two phases. The second phase consists of an upgrade with a larger and more powerful magnet—which has also been used before in MRI. DALI is currently in the design and prototyping stage. 

{As shown in Fig. \ref{Fig_2} and Fig. \ref{Fig_3}, the numerical simulation which has been developed in this work indicates that the first phase of DALI project is capable of detecting a DFSZ I axion at $\sim$$ 4\sigma$ with an exposure time of about two weeks per 50 MHz-wide spectrum. An upgrade with a larger 11.7 T magnet is planned which will greatly accelerate the scanning speed, shortening the acquisition time needed to probe a DFSZ I axion with a $\sim$5$\sigma$ significance to about one day. This represents an reduction of the time needed to probe a given mass range with a certain confidence level by a factor of $\gtrsim$20.}

In the Monte Carlo simulation reported in this work, artificial axion signals have been injected at a frequency close to the resonator quality factor peak. As the axion signal is placed further away, its SNR decreases in proportion to $Q$. During the calibration procedure, a measurement of both white noise and 1/f noise is expected to be performed with the goal of measuring the knee frequency of the receiver \cite{NRAO_Radio_Telescopes_and_Radiometers}. A lower value of f$_{\mathrm{knee}}$, which is of the order of tens of hertz, could allow for a higher frequency resolution. In addition, results presented in this manuscript have assumed a realistic noise level, whose magnitude and spectral characteristics may however be different in the final setup and therefore alter the sensitivity of the instrument—e.g. higher or lower physical temperatures, systematics. Finally, persistent candidates are to be re-scanned by ramping-down the superconducting magnet that are analyzed in parallel. 

In conclusion, the results of the Monte Carlo simulation reported in this manuscript suggest that sensitivity of the DALI Experiment to QCD-inspired axion models is tenable within a mass range that remains under-examined due to the hurdles associated with the development of the experimental approach requiered to explore it.

\vspace{6pt} 

\textbf{Acknowledgements:} We thank R. Hoyland, R. Rebolo, H. Lorenzo-Hernández for discussions. We thankfully acknowledge the technical expertise and assistance provided by the Spanish Supercomputing Network (Red Española de Supercomputación), as well as the computer resources used: the deimos-diva supercomputer, located at the Instituto de Astrofísica de Canarias.

\end{document}